\documentclass[10pt,twocolumn,english,pre]{revtex4}

\usepackage{amsmath}    
\usepackage{graphicx}   

\usepackage[T1]{fontenc}
\usepackage[latin1]{inputenc}
\usepackage{graphicx,epsfig}
\usepackage{html}
\usepackage{natbib}

\begin{document}
\def\deg{^\circ}
\def\AA{{\rm A}}
\def\half{ \small {1 \over 2}}

\title{The Boltzmann factor, DNA melting, and Brownian ratchets:
Topics in an introductory physics sequence for biology and premedical students}

\author{S. G. J. Mochrie}
\affiliation{Departments of Physics and Applied Physics, Yale University, New Haven, Connecticut 06511}

\date{\today}

\pacs{pacs}

\begin{abstract}
Three, interrelated biologically-relevant  examples of biased random walks are presented:
(1) A model
for DNA melting, modelled as DNA unzipping,
which provides a way to illustrate
the role of the Boltzmann factor in a venue well-known to biology
and pre-medical students;
(2)  the activity of helicase motor proteins in unzipping double-stranded DNA,
for example, at the replication fork,
which is an example of a Brownian ratchet;
(3) force generation by actin polymerization, which
is another  Brownian ratchet, and for which
the force and actin-concentration dependence of  the velocity of actin polymerization
is determined.
\end{abstract}

\maketitle

\section{Introduction}

In spite of the growing recognition that physics skills -- ``scholastic rigor, analytical thinking, quantitative assessment, and the analysis of complex systems'' 
\cite{AAMC} -- are important for biology \cite{BIO2010}
and pre-medical \cite{AAMC} students,
these students often arrive in
physics classes
skeptical about the relevance of physics to their academic and
professional goals.
To engage these students, in the 2010-2011 academic year, the Yale physics department debuted a new introductory physics sequence,
that,  in addition to covering the basics
-- kinematics, force, energy, momentum, Hooke's Law, Ohm's Law, Maxwell's equations {\em etc.} -- also covers a number of more biologically-relevant topics, including,
in particular, probability, random walks, and the Boltzmann factor.
The point of view of the class is that the essential aspect of physics is that it constitutes a mathematical description of the natural world, irrespective of whether the topic
is planetary motion or cellular motion.

The enrollment in the new sequence was approximately 100 students.
The class is evenly split between sophomores and juniors with a few seniors.
The majority (80\%)  are biology majors, with 80\% 
identifying themselves as premedical students,  and they possess considerable biological
sophistication.
In many cases, they are involved in biomedical research at Yale or at the
Yale School of Medicine. In many cases too, they are involved in medically-related volunteer work.
The major time commitment required to do justice to a rigorous physics class has to compete with these other obligations. Therefore, an important aspect of our teaching strategy is to convince these students
that physics  is indeed relevant to their goals.
To this end, we determined to cover a number of biologically-relevant topics,
with which the majority of the students would have some familiarity
from their earlier biology and chemistry classes.

This paper presents three such topics,
that are interrelated and can be treated as random walks, in the
hope that these  may be useful to others.
First is
DNA melting \cite{Kittel1969},
which we place in the context of  polymerase chain reaction (PCR).
This provides a way to illustrate
the role of the Boltzmann factor in a venue well-known to the students. 
This treatment builds on earlier sections of the course,
concerned with random walks and chemical reaction rates,
which are not described here.
The second topic is
the activity of helicase motor proteins in unzipping double-stranded nucleic acid (DNA or RNA,
although we will write in terms of DNA).
Our discussion is based on Ref. \onlinecite{PhysRevLett.91.258103}.
Helicase activity constitutes an elegant example of a Brownian ratchet
and  builds on the earlier discussion of DNA melting.
Third, we present a discussion of force generation by actin polymerization,
which provides the physical basis of cell motility in many cases, and which
is another  Brownian ratchet.
In this case, based on Ref. \onlinecite{Peskin1993},
we can determine how the velocity of actin polymerization depends on actin
concentration and on load.
In each of these examples,
biology and pre-medical
students in an introductory physics class see that a physics-based approach permits a
new, deeper understanding of a familiar
molecular-biological phenomenon.

\section{The Boltzmann Factor}
\noindent
"The laws of thermodynamics may easily be obtained from the principles of
statistical mechanics,
of which they are an incomplete expression." 
J.W. Gibbs \cite{Gibbs}.

Instead of  introducing thermal phenomena via thermodynamics and heat engines, as might occur in a traditional introductory
sequence,  following  the suggestion of Garcia {\em et al.} \cite{Garcia2007b}, we chose to 
assert the Boltzmann factor as the fundamental axiom of
thermal physics.
Building upon earlier
sections of the course on probability and random walks, this approach
permits us to rapidly progress to physics-based treatments of DNA melting,
unzipping of double-stranded DNA at the replication fork
by helicase motor proteins,  and force-generation by actin-polymerization.
Specifically, we assert that,
for microstates $i$ and $j$ of a system,
the probability ($p_i$) of realizing a  microstate $i$  and
the probability ($p_j$) of realizing a  microstate $j$
are related via
\begin{equation}
\frac{p_i}{p_j} = e^{-(\epsilon_i-\epsilon_j)/(k_B T)},
\label{BF}
\end{equation}
where
 $\epsilon_i$ is the energy of microstate $i$,
  $\epsilon_j$ is the energy of microstate $j$,
 $k_B = 1.38 \times 10^{-23}$~JK$^{-1}$ is Boltzmann's constant, and  $T$ is the
 absolute temperature.

\vspace{0.1in}
\noindent
"This fundamental law is the summit of statistical mechanics, and the entire subject is either the slide-down
from this summit, as the principle is applied to various cases, or the climb up to
where the fundamental law is derived and the concepts
of thermal equilibrium and temperature clarified."
R. P. Feynman on the Boltzmann factor \cite{Feynman}.
\vspace{0.1in}

To illustrate the Boltzmann factor in a simple example,
we consider protein folding/unfolding.
 Protein/unfolding is  an example of an isomerization reaction,
 in which one chemical species alternates
between different molecular configurations.
In this case, it is important to
realize that the folded state corresponds to a single microstate, but that the
unfolded state corresponds to $g$ microstates.
This is because there is just one molecular configuration
associated with the folded state. By contrast, the unfolded state can be viewed as
a random walk in space, and therefore corresponds to $g$ different molecular
configurations,
one for each different random walk.
If there are a total of $n$ proteins, $n_u$ of which are unfolded, and if there are $g$ possible unfolded microstates, then the probability of realizing a particular unfolded microstate ($p_u$) is
equal to the probability
that a protein molecule is unfolded multipled by the probability that an unfolded protein is
in the particular unfolded microstate of interest,
which is one of $g$ equally-likely microstates:
\begin{equation}
p_u=\frac{n_u}{n} \times \frac{1}{g}.
\label{EQ2}
\end{equation}
There is a unique folded microstate, so in terms of $n$ and the number of folded proteins,
$n_f$, the
probability of realizing the folded microstate is simply
\begin{equation}
p_f =  \frac{n_f}{n},
\label{EQ3}
\end{equation}
Combining  EQ. \ref{BF}, EQ. \ref{EQ2}, and EQ. \ref{EQ3}, we find
\begin{equation}
\frac{n_u}{n_f} =g e^{-(\epsilon_u - \epsilon_f)/(k_B T)},
\label{xxx}
\end{equation}
where $\epsilon_u$ is the energy of any of the unfolded states
and $\epsilon_f$ is
the energy of the folded state.

\section{DNA melting}
\subsection{DNA unzipping/zipping as a chemical reaction}
\begin{figure}[t!]
\begin{center}
{\includegraphics[width=0.47\textwidth,keepaspectratio=true]{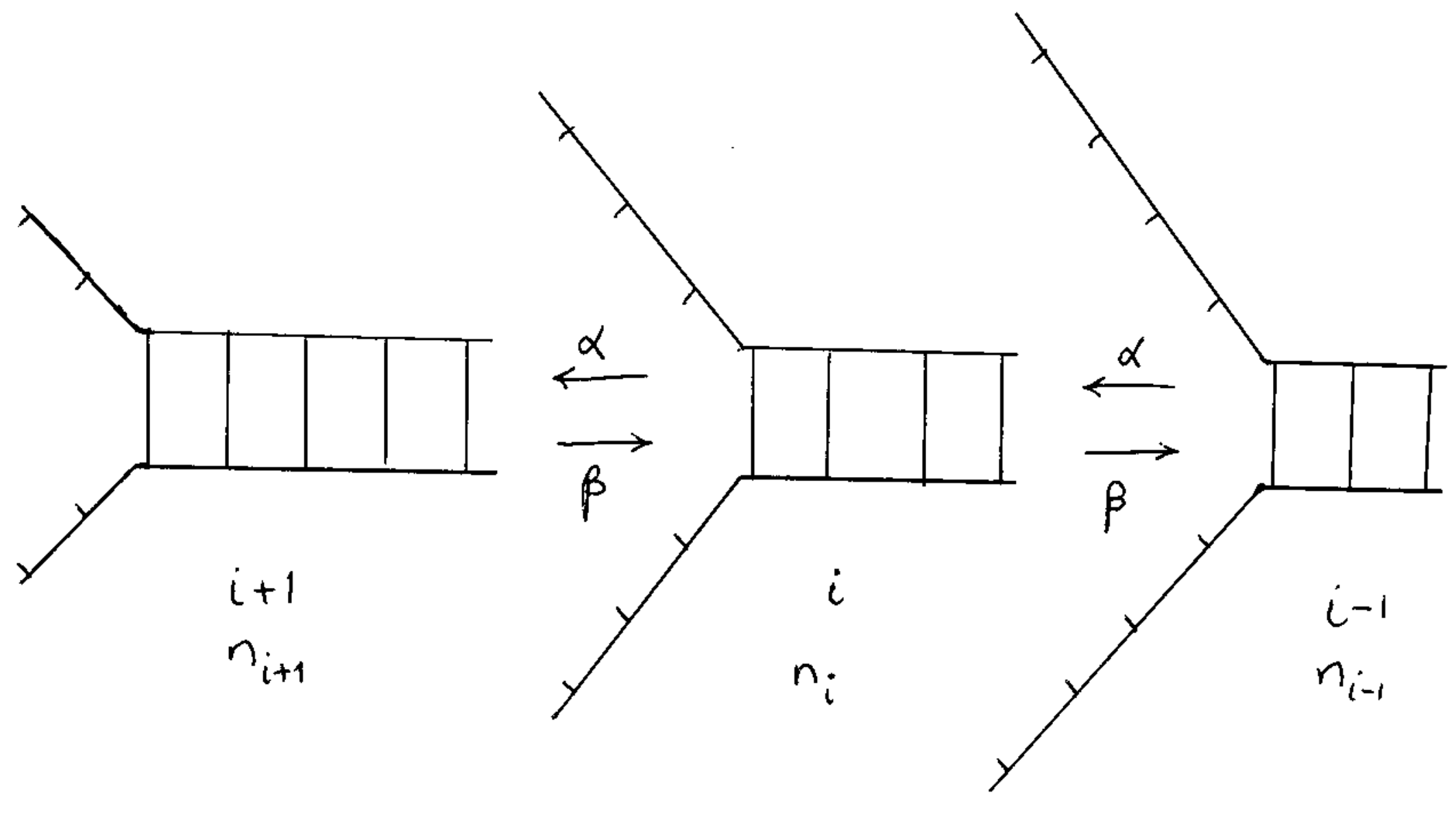}}
\end{center}
\caption{Chemical reaction scheme for DNA zipping and unzipping.
The possible reactions of a DNA strand with $i$ zipped base pairs
are illustrated, either undergoing isomerization to a DNA strand with
${i-1}$ base pairs or isomerization to a DNA strand with ${i+1}$ base pairs.}
\label{FIG1}
\end{figure}

Next, we examine  DNA melting, according to the
model of Ref. \onlinecite{Kittel1969}, in which DNA melting is
equivalent to DNA unzipping.
We treat DNA zipping and unzipping as a set of isomerization reactions. 
To this end, we consider a population of identical
DNA strands each of which contains a junction between dsDNA and ssDNA.
Fig. \ref{FIG1} illustrates the reactions involving the DNA strand with $i$ paired base pairs. This is the
chemical species in the center. The species on the left and right are DNA strands with $i+1$ and $i-1$
paired base pairs, respectively.
The relevant reaction rates are
$\alpha$, which is the zipping rate, and $\beta$, which is the unzipping rate.
When $\alpha > \beta$, the DNA zips up. When $\alpha < \beta$, the DNA unzips.
As suggested in Fig. \ref{FIG1},
$n_i$ is the {mean} number of DNA strands with $i$ paired base pairs, {\em etc.}

We have previously discussed in class that how the concentration of chemical species changes in time can be described
by chemical rate equations.
With the help of Fig. \ref{FIG1}, we are thus lead to
an equation for the rate of change of $n_i$ in terms of $\alpha$, $\beta$, $n_i$, $n_{i-1}$, and
$n_{i+1}$:
\begin{equation}
\frac{dn_i}{dt} = -\alpha n_i - \beta n_i + \alpha n_{i-1} + \beta n_{i+1}.
\end{equation}
At equilibrium, at a temperature $T$,  on-average nothing changes as a function of time, so $dn_i/dt =0$.
Thus, 
\begin{equation}
0 = -\alpha - \beta + \alpha \frac{n_{i-1}}{n_i} + \beta \frac{n_{i+1}}{n_i}.
\label{SteadyState2}
\end{equation}

\vspace{0.1in}
\noindent
The factor $n_{i-1}/n_i$,
which is the ratio of  the mean number of DNA strands with $i-1$ zipped base pairs to the mean number
with $i$ zipped base pairs, is equal to the ratio of the
probability that a particular DNA strand has $i-1$ zipped base pairs to the probability that it has
$i$ zipped base pairs. Thus, this factor is given by a Boltzmann factor ({\em cf.} EQ. \ref{xxx}):
\begin{equation}
\frac{n_{i-1}}{n_i} = g e^{-\epsilon/(k_B T)} = e^{-(\epsilon - k_B T \ln g)/(k_B T)} = e^{-\Delta G/(k_B T)},
\label{zip1}
\end{equation}
where $\epsilon$ is the energy required to unzip one additional base pair
(so $\epsilon$ is positive) and $g$ specifies that the two unzipped ssDNA bases have a factor
$g$ times as many microstates as the single dsDNA base pair they replace.
The last equality in EQ. \ref{zip1} defines
the free energy required to unzip one base pair:
\begin{equation}
\Delta G = \epsilon - k_B T \ln g.
\label{EQ4}
\end{equation}
Students are familar with $\Delta G$ from their chemistry classes.
Similarly, we have
\begin{equation}
\frac{n_{i+1}}{n_i}  = e^{+\Delta G/(k_B T)}.
\label{zip2}
\end{equation}

\vspace{0.1in}
\noindent
Substituting EQ. \ref{zip1} and EQ. \ref{zip2} into EQ. \ref{SteadyState2}, we have
\begin{equation}
0 = -\alpha - \beta + \alpha e^{-\Delta G/(k_B T)} + \beta e^{+\Delta G/(k_B T)}.
\label{SteadyState3}
\end{equation}
It follows from EQ. \ref{SteadyState3} that
\begin{equation}
\frac{\beta}{\alpha} = e^{-\Delta G/(k_B T)}.
\label{DG}
\end{equation}
EQ. \ref{DG} informs us that the DNA unzips, {\em i.e.}  $\beta >\alpha$, only if $\Delta G < 0$,
{\em i.e.} only if $\epsilon - k_B T \ln g < 0$. In order for this condition  ($\epsilon - k_B T \ln g < 0$) to be
satisfied, it is necessary that $T > \epsilon / (k_B \ln g)$. If we define the DNA ``melting temperature'' to be
$T_M = \epsilon / (k_B \ln g)$, we see that the DNA unzips for $T > T_M$, while it zips up for $T<T_M$.
This phenomenon is an essential ingredient in DNA multiplication by polymerase chain reaction (PCR),
which is well-known to the students, and for which
Kary Mullis won the 1993 Nobel Prize in Chemistry  \cite{PCR}.
The first step in PCR is to raise the temperature
above $T_M$, so that each dsDNA strand unzips to become two ssDNA strands.
When the temperature is subsequently reduced in the presence of oligonucleotide primers,
nucleotides and DNA polymerase,
each previously-unzipped, ssDNA strand templates its
own conversion to dsDNA. This doubles the original number of dsDNA strands because  a
new dsDNA strand is created for each  ssDNA.
PCR involves
repeating this temperature cycling process multiple ($N$)  times, with the result that the initial
number of dsDNA molecules is multiplied by a factor of $2^N$. 
Thus, initially tiny quantities of dsDNA can be hugely amplified,
and
subsequently  sequenced.

\subsection{DNA unzipping/zipping from a random-walk point of view}
It is also instructive to
view DNA zipping/unzipping as a biased random walk,
which students have previously studied in the class.
In this context, if we consider a dsDNA-ssDNA junction,
the probability of zipping up one base pair in a time $\Delta t$ is $\alpha \Delta t$, and
the probability of unzipping one base pair in a time $\Delta t$ is $\beta \Delta t$.
For small enough $\Delta t$
it is reasonable to assume that the only
three possibilities are (1) to zip up one base pair or (2)  to unzip one base pair or (3) to not do
anything.
Therefore, since probabilities sum to unity, we must have that the probability
to do nothing is $1 - \alpha \Delta t - \beta \Delta t$.
Given these probabilities, and the length of a base pair, $b$,
we may readily calculate the mean displacement of the ssDNA-dsDNA junction in a time $\Delta t$:
\begin{equation}
\Delta x_j = 
b(\beta-\alpha) \Delta t,
\label{DxJ}
\end{equation}
where zipping corresponds to a negative displacement of the ss-to-ds junction
Since the mean of the sum of $n$ identically-distributed, statistically-independent random variables
is $n$ times the mean of one of them (which students  learned earlier in the course),
then in a time $t = n \Delta t$ the mean displacement of the ssDNA-dsDNA
junction is
\begin{equation}
x_j = n \Delta x_j = \frac{t}{\Delta t} \Delta x_j =  \frac{t}{\Delta t} b(\beta-\alpha) \Delta t = b(\beta-\alpha) t.
\end{equation}
The corresponding drift velocity of the ssDNA-dsDNA junction is
\begin{equation}
v_j =  b(\beta-\alpha) = b \alpha(e^{-\Delta G/(k_BT)}-1)
\label{vJ}
\end{equation}
This is the drift velocity of a dsDNA-ssDNA junction in terms of the zipping up rate ($\alpha$) and the unzipping
rate ($\beta$), or the zipping rate ($\alpha$) and the unzipping free energy ($\Delta G$).
We will come back to this result below, but we note now that EQ. \ref{vJ} is appropriate { only}
when the junction is far from a helicase.

As defined in EQ. \ref{EQ4},
$\Delta G$ is the change in free energy that occurs when one additional base pair is
unzipped.
Thus, as far as this expression for $\Delta G$ is concerned,
the final,  ``product'' state is the unzipped state, and the initial,  ``reactant''
state is the zipped state. Thus, unzipping corresponds to the forward direction of the reaction.
We may make contact with what students have learned in chemistry classes, namely that a reaction proceeds forward if
$\Delta G$ is negative, by pointing out that
EQ. \ref{vJ} informs us that the unzipping reaction proceeds forwards
({\em i.e.} that $v_j>0$) only  for $\Delta G < 0$, exactly as we
are told in chemistry classes. Here, though, this result is derived from a more
basic principle, namely the Boltzmann factor.

\section{Helicase DNA-unzipping activity}
\label{sec}
\subsection{Helicases}
Helicases \cite{Pyle2008}
are a class of motor proteins (a.k.a.
molecular motors),  which perform myriad tasks in the
cell by catalyzing ATP-to-ADP hydrolysis and using the free energy released
in this reaction to do work.
The importance of helicases may be judged from the fact that 4\% of the
yeast genome codes for some kind of helicase.
One of their roles is to unzip dsDNA and/or dsRNA.
Thus, helicases play an indispensible role in
DNA replication, for example.
To engage  students in this topic, we start by showing a number of
online movies illustrating the DNA-unzipping activity of
helicase motor proteins at the replication fork
\cite{movie1,movie2}.
These movies also
present an opportunity for active
learning in which we ask students to discuss with their neighbors
what is misleading about the videos.
The essential point is that, wonderful as they are, the videos suggest
that everything proceeds deterministically.
By contrast, as we will discuss, all of the processes depicted are actually random
walks, but with a drift velocity that corresponds to their progress.
We also  point out that, on the medical side, Werner syndrome,
which involves accelerated aging, is caused by a mutation in the
{\em WRN} gene which codes for the helicase WRN \cite{Werner}.

\subsection{Brownian ratchet mechanism of helicase activity}
One proposed mechanism for how helicase unzips DNA  is as follows. The helicase steps unimpeded
on ssDNA towards a ss-to-ds junction, until it encounters the junction, which then
blocks its further progress, because the helicase translocates only on ssDNA.
However, at the junction,
there is a non-zero
probability per unit time for the junction to thermally unzip one base pair, because of the Boltzmann factor.
It is then  possible
for the helicase to step into the just-unzipped position.
If the helicase does this,  the DNA is prevented from subsequently zipping back
up again. In this way, the junction is unzipped one step.
Repeating this process many times leads to the complete unzipping of the DNA.
Because this mechanism relies on random Brownian motions
to both unzip the DNA and to move
the helicase into the just-unzipped position, the helicase is said to be a {Brownian ratchet}
\cite{Peskin1993},
analogous to  Feynman's
thermal ratchet \cite{FeynmanLectures}.

\subsection{Helicase translocation from a random-walk point of view}
Just as the motion of the ss-to-ds junction may be conceived as a random walk, so may
be the translocation of the helicase on ssDNA.
In this case,
the probability of the helicase stepping one base pair towards the junction ($+b$) in a time $\Delta t$ is $k_+ \Delta t$, and
the probability stepping one base pair away from the junction ($-b$) in a time $\Delta t$ is $k_- \Delta t$,
where $k_+$ and $k_-$ are the rate of stepping towards the junction and the rate of
stepping away from the junction, respectively. Since probabilities sum to unity, and we assume that the only
three possibilities in a small time $\Delta t$ are to step towards the junction one base pair or to
step away from the junction one base pair or to not do
anything, we must have that the probability to do nothing is $1 - k_+ \Delta t - k_- \Delta t$.
Given these probabilities, and the length of a base pair, $b$,
we may calculate the mean displacement of the helicase in a time $\Delta t$:
\begin{equation}
\Delta x_h =
b(k_+ - k_-) \Delta t.
\label{DxH}
\end{equation}
Since the mean of the sum of $n$ identically-distributed, statistically-independent random variables
is $n$ times the mean of one of them, then in a time $t = n \Delta t$ the mean displacement of the helicase is
\begin{equation}
x_h = n \Delta x_h = \frac{t}{\Delta t} \Delta x_h =  \frac{t}{\Delta t} b(k_+ -k_-) \Delta t = b(k_+ - k_-) t.
\end{equation}
The corresponding drift velocity of the helicase is
\begin{equation}
v_h  =  b(k_+ - k_-).
\label{vH}
\end{equation}
This is the drift velocity of a helicase in terms of the stepping-towards-the-junction rate ($k_+$) and the
stepping-away-from-the-junction rate ($k_-$). 
Just like EQ. \ref{vJ}, EQ. \ref{vH}  is appropriate { only}
when the helicase is far from the junction.
\begin{figure}[t!]
\begin{center}
{\includegraphics[width=0.45\textwidth,keepaspectratio=true]{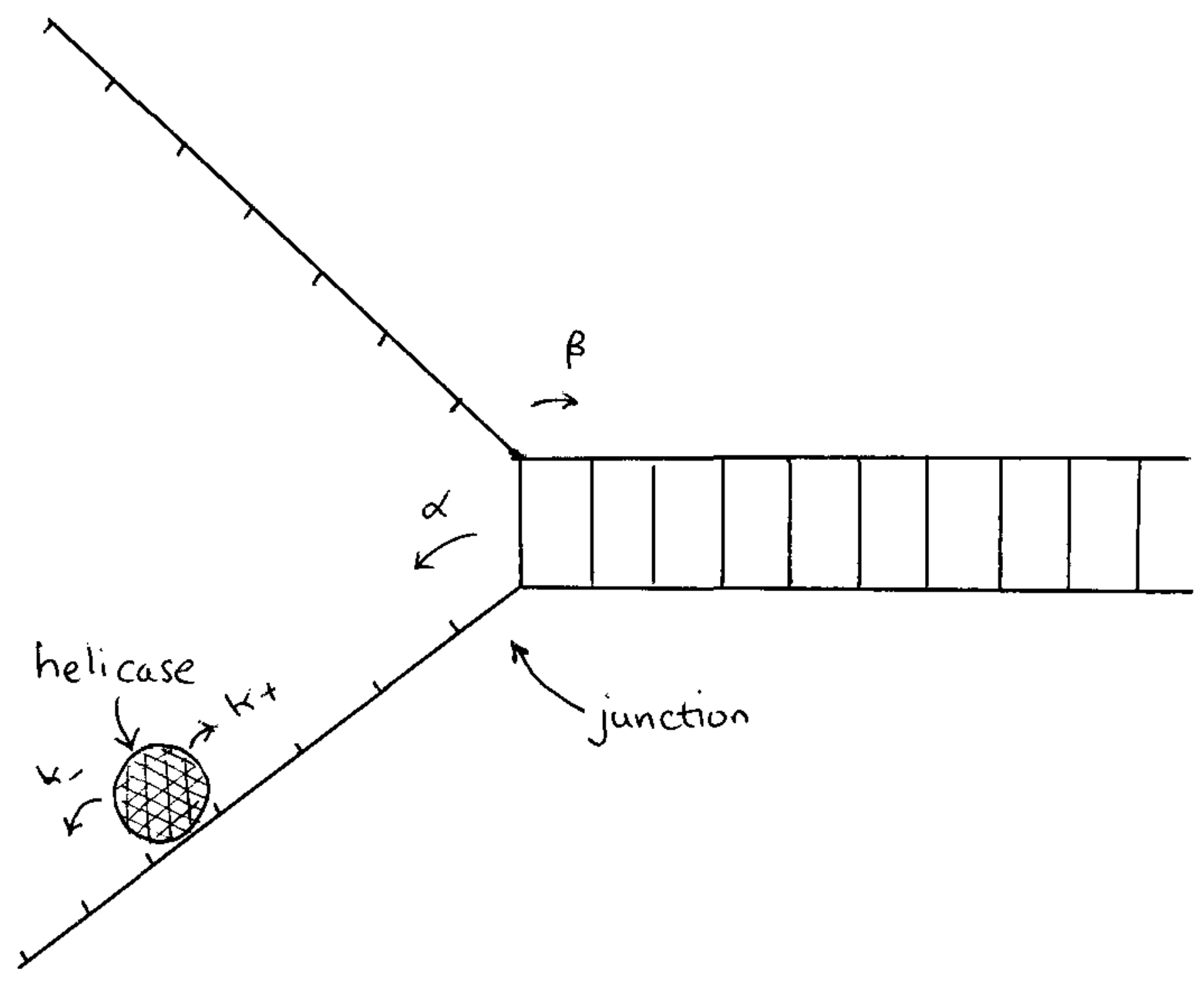}}
\end{center}
\vspace{-0.1in}
\caption{Schematic of helicase translocation on ssDNA and
DNA zipping, showing the relevant parameters. In the context of this figure,
the coordinate system used in our discussion takes the $x$-direction to increase towards the right,
so that $\alpha$ and $k_-$ correspond to motion in the negative $x$-direction, and
$\beta$ and $k_+$ correspond to motion in the positive $x$-direction.}
\label{HJ2}
\vspace{0.in}
\end{figure}

An important additional point, concerning helicase translocation on ssDNA, is that,
as we saw in EQ. \ref{DG}, the ratio of forward and backward rates is given by a change in free energy.
Thus, for helicase stepping we must expect, in analogy with EQ. \ref{DG}, that the ratio of
stepping rates is given by
\begin{equation}
\frac{k_+}{k_-} = e^{\Delta G'/(k_B T)},
\label{DGprime}
\end{equation}
where $\Delta G'$ is a free energy change.
But what free energy change? The answer can be gleaned from the observation that
helicases, and motor proteins generally, can be thought of as enzymes, which catalyze
ATP-to-ADP hydrolysis, which is coupled to the helicase's translocation.
It follows that $\Delta G'$ in EQ.  \ref{DGprime} corresponds to the free energy difference between
ATP and ADP. (Note that $\Delta G'$, as specified in EQ. \ref{DGprime}, must be positive, in
order to ensure that $k_+ > k_-$ so that the helicase
translocates on ssDNA preferentially towards the ssDNA-to-dsDNA junction.)

\subsection{Clash of the titans}
\begin{figure}[t!]
\begin{center}
{\includegraphics[width=0.47\textwidth,keepaspectratio=true]{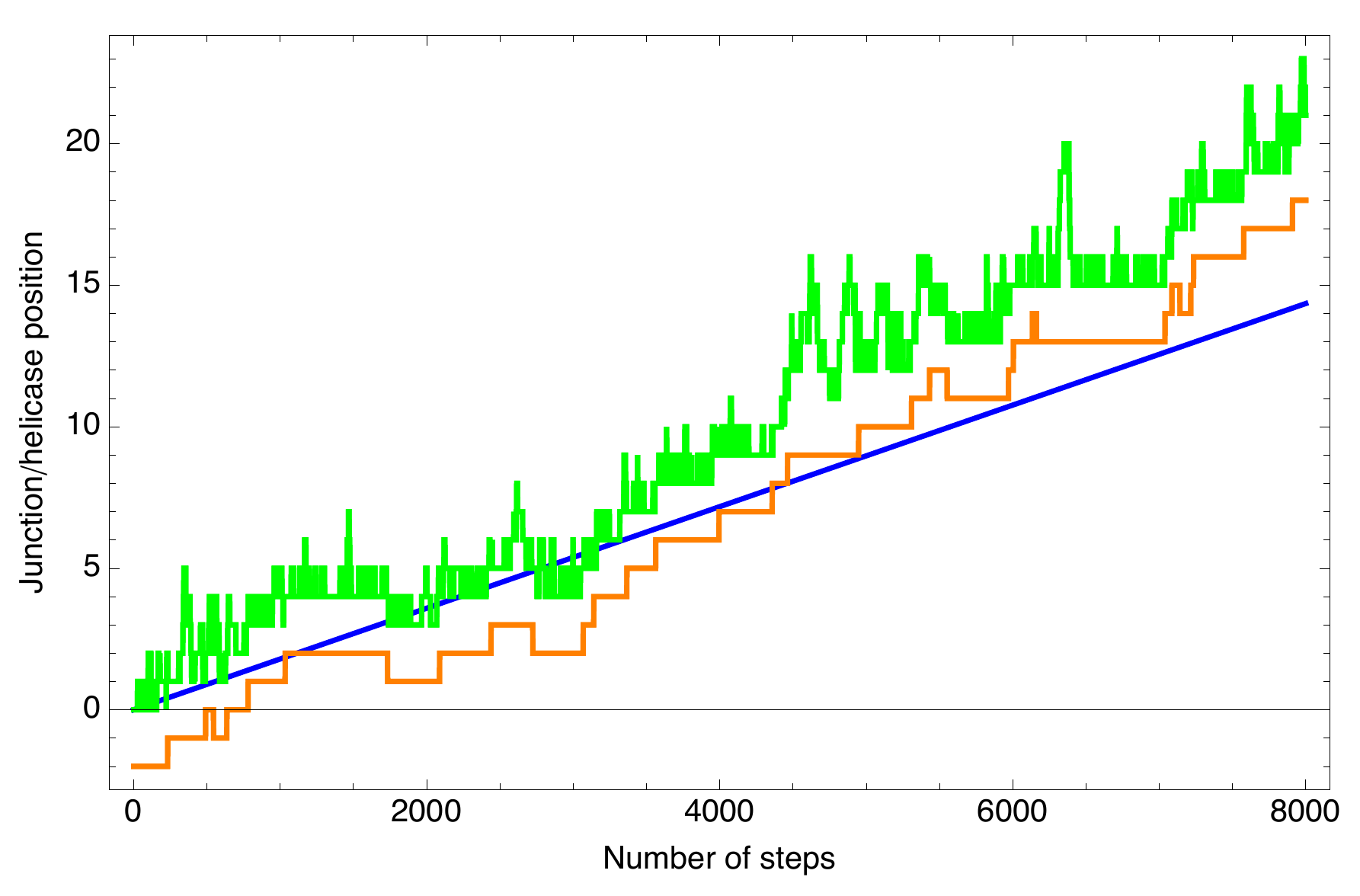}}
\end{center}
\vspace{-0.1in}
\caption{
Simulation, implemented as a Mathematica Demonstration,
of a helicase motor protein unzipping double-stranded DNA, according
to the Brownian ratchet model of helicase activity.
The orange trace represents  helicase
translocation on ssDNA (left) for $k_+ = 0.79$ and $k_-=0.07$.
The green random walk represents the position of a ssDNA-to-dsDNA junction for
$\alpha = 7.2$ and $\beta = 2.45$. In both cases, $\Delta t = 0.01$.
The blue line is EQ. \ref{vvv}.
The two random walks start 
at 0 in the case of the junction, and at -2 in the case of the helicase.
}
\label{TITANS}
\end{figure}

So far, we have considered the situtation  when the ds-to-ss junction and the helicase are
far apart. To determine how helicase unzips dsDNA, it is necessary to determine what happens when these
two objects come into close proximity, given that they
cannot cross each other.
To elucidate what happens in this case, we show to the class a simple
Mathematic Demonstration
that simulates these two non-crossing
random walks \cite{demo}.
The simulation treats both the location of the helicase and the location of the
ds-to-ss junction as random walks. At each time step within the simulation, the helicase ordinarily
steps in the positive $x$-direction,
towards the junction, with probability $k_+ \Delta t$ and in the negative $x$-direction, away
from the junction, with probability $k_-\Delta t$,
while the junction ordinarily steps in the positive $x$-direction,
zipping up one step, with probability $\alpha \Delta t$
and in the negative $x$-direction, unzipping one step, with probability $\beta \Delta t$.
However, in the simulation, if the helicase and the junction are neighbors, neither one
is permitted to step to where it would overlap with the other. Thus, the helicase and the junction
cannot cross.
An example of the simulational results is shown in Fig. \ref{TITANS},
where
the orange trace represents
helicase location as a function of time and the green trace represents the location of the ssDNA-to-dsDNA junction as a function of time.
Evidently,
the helicase and the junction track together, implying that they have the same
drift velocity.
For the parameters of this simulation, the helicase translocates in
the same direction as it would in the absence of the junction. By contrast, the junction's
direction is opposite its direction without the helicase.
Thus, for these parameters, the helicase indeed unzips dsDNA.
Using sliders within the Mathematica Demonstration, which is readily accessed
via any web browser, students can
explore for themselves the effects of varying $\alpha$, $\beta$, $k_+$, and $k_-$.

To incorporate analytically the fact that the helicase
 and the junction can not cross,
we introduce the probability, $P$, that the helicase and the junction are not next to each other.
The dsDNA-to-ssDNA junction can only zip up if the helicase and the junction are not next to each other. 
Therefore, we reason that EQ. \ref{DxJ} should be modified to read
\begin{equation}
\Delta x_j  =  b(\beta -\alpha P) \Delta t.
\end{equation}
Similarly, EQ. \ref{DxH} should be modified to read
\begin{equation}
\Delta x_h  =  b(k_+P  - k_- ) \Delta t.
\end{equation}
It follows that the drift velocities are modified to read
\begin{equation}
v_j  =  b(\beta-\alpha P),
\label{vJ2}
\end{equation}
and
\begin{equation}
v_h  =  b(k_+ P -k_-).
\label{vH2}
\end{equation}. 

However, from the simulation, it is also clear that,
while the helicase is unzipping DNA, the helicase and the junction must have the same drift velocity {\em i.e.}
\begin{equation}
v_h  =  v_j
\end{equation}
or
\begin{equation}
b(k_+ P-k_- ) = b(\beta-\alpha P).
\end{equation}
We can solve this equation to determine $P$:
\begin{equation}
P = \frac{k_- + \beta}{k_+ +  \alpha}.
\label{P}
\end{equation}
Furthermore, we can use this expression for $P$
to determine the drift velocity at which the
helicase unzips the dsDNA by substituting
into EQ. \ref{vH2}. Setting $v_j=v_h = v$, we find
\begin{equation}
v  =  b \left ( \frac{\frac{\beta}{\alpha}-\frac{k_-}{k_+}   }{\frac{1}{\alpha} +  \frac{1}{k_+ }} \right ).
\label{vvv}
\end{equation}
EQ. \ref{vvv} represents the velocity at which helicase unzips dsDNA according to the
Brownian ratchet mechanism.
The numerator in EQ. \ref{vvv} is the difference of two rate ratios.
It follows,
using  EQ. \ref{DG}
and EQ. \ref{DGprime} in EQ. \ref{vvv}, that
\begin{equation}
v= b \left ( \frac{ e^{-\Delta G/(k_BT)}-e^{-\Delta G'/(k_BT)}}{\frac{1}{k_+} + \frac{1}{ \alpha} } \right ),
\label{vvvv}
\end{equation}
EQ. \ref{vvvv} informs us that
whether or not helicase unzips dsDNA  depends solely on
whether  $\Delta G' > \Delta G$ or not.
For $\Delta G' > \Delta G$,  the drift velocity of the helicase-plus-junction is positive, corresponding to
the helicase unzipping the dsDNA.
In fact, for one base pair, we have $\Delta G \simeq 3 k_B T$, while for the
hydrolysis of one ATP molecule, we have $\Delta G' \simeq 16 k_B T$,
so indeed the helicase has plenty of free energy to
do its work. In fact, energetically, one ATP hydrolysis cycle could unzip up to about 5 base pairs.

In fact, beautiful, single-helicase experiments \cite{Johnson2007}
suggest that the simple Brownian ratchet mechanism
of helicase DNA-unzipping activity, presented here,
 should be refined by incorporating both a softer repulsive potential
between the helicase and the ds-to-ss junction than the hard-wall potential implicit in our discussion,
and suitable free energy barriers between different microstates of the helicase and junction
\cite{PhysRevLett.91.258103,Johnson2007,Pyle2008}. 
Appropriate choices of the potential and the barriers
permit the helicase to unzip dsDNA faster than would occur
in the case of a hard-wall potential.
A force ($f$), that tends to unzip the DNA, can be
incorporated by replacing $\alpha$ with $\alpha e^{-f b/(k_B T)}$ and $\beta$
with  $\beta e^{f b/(k_B T)}$. 

\section{Force generation by actin polymerization}

\subsection{Actin polymerization is a Brownian ratchet}
\begin{figure}[t!]
\begin{center}
\includegraphics[width=0.49\textwidth,keepaspectratio=true]{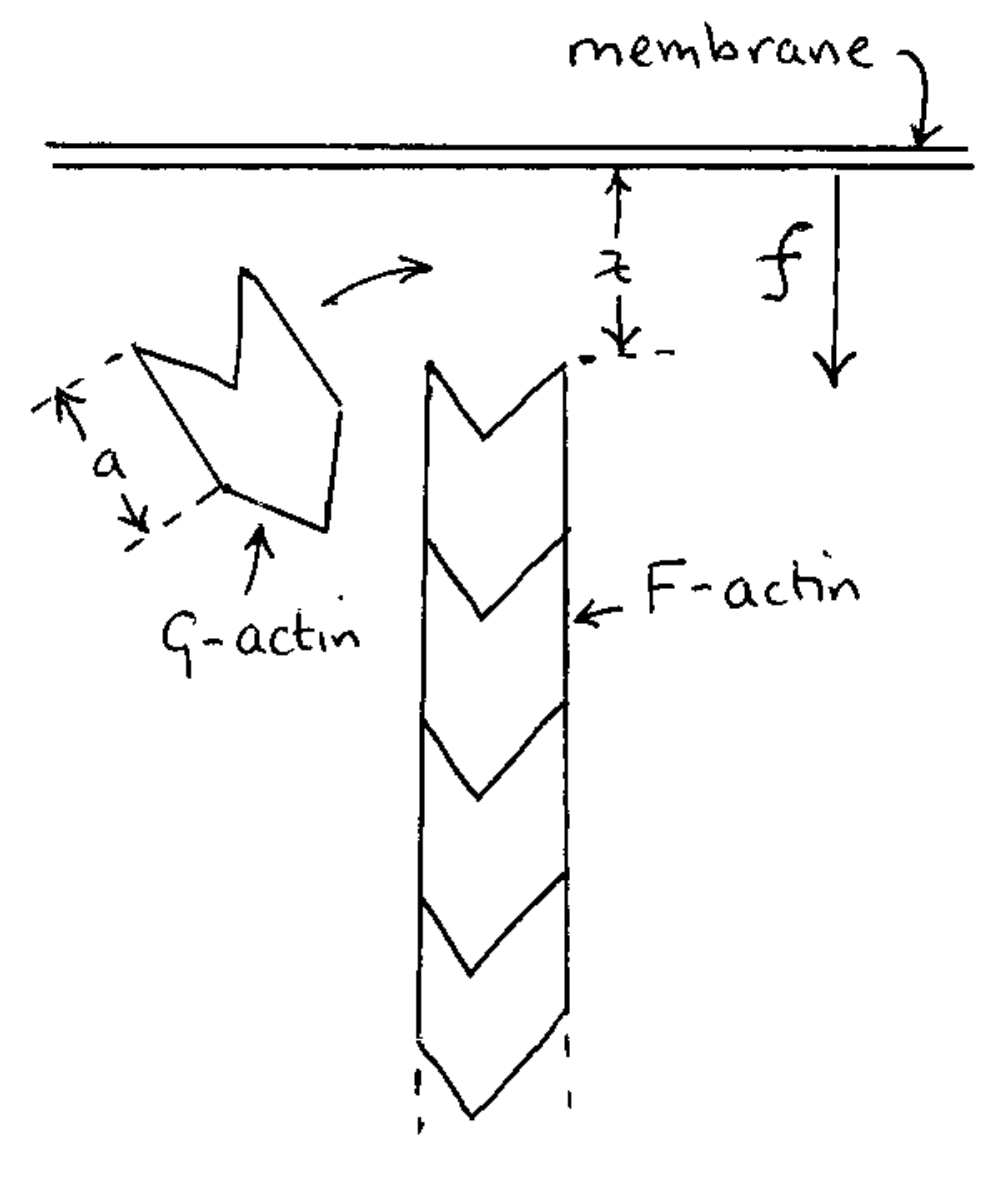}
\end{center}
\caption{
Cartoon illustrating how actin polymerization can do work against a load ($f$), applied to a membrane
against which the polymerizing actin filament (F-actin) abuts. Only if the gap, $z$, between the tip of the
actin filament and the membrane exceeds the length, $a$, of a G-actin monomer is it possible for the
filament to grow.
}
\label{Actin}
\end{figure}

The mechanism by which actin or tubulin polymerization exerts a force 
also a Brownian ratchet \cite{Peskin1993,Dogterom1997,Kovar2004} and is schematically illustrated in
Fig. \ref{Actin}.
In the case of  a load, $f$, applied to a cell membrane, the cell membrane is in turn
pushed against the tip of an actin filament, which usually
prevents the addition of an additional actin monomer (G-actin) of  length $a$ to the tip of the actin
filament (F-actin).
However, with probability specified by a Boltzmann factor, the membrane's position relative
to the tip, $z$, occasionally
fluctuates far enough away from the filament tip ($z>a$)
to allow a monomer to fit into the gap.
If a  monomer does indeed insert and add to the end of the  filament,
the result is that the filament and therefore the membrane move one step forward,
doing work against the load force.
Repeating this many times for many such filaments gives rise to cell motility against viscous forces.
In class we also show movies showing cells moving as a result of actin polymerization
\cite{rocket1},
and {\em Listeria monocytogenes} actin ``rockets''
\cite{rocket}.

\subsection{Actin polymerization as a biased random walk}
Similarly to EQ. \ref{DxJ} and EQ. \ref{DxH}, we can write down an expression for the
mean
displacement of the filament tip in a time $\Delta t$
in the absence of a nearby membrane:
\begin{equation}
\Delta x =a(c k_+ - k_-) \Delta t
\label{actinX}
\end{equation}
where $c$ is the concentration of G-actin, $a$ is the length of an actin monomer, $k_+$ is the actin on-rate, and $k_-$ is the actin off-rate.
However, if the membrane is nearby, it is only possible to add an actin monomer if the distance
between the filament and the membrane is greater than $a$.
Assuming that the time-scale for membrane fluctuations is much faster than that for
adding actin monomers, if the probability, that the
membrane-filament tip distance is greater than $a$, is $P$, then EQ. \ref{actinX} is modified to read
\begin{equation}
\Delta x =a(c k_+ P - k_-) \Delta t,
\label{actinX}
\end{equation}
and the drift velocity of the tip is
\begin{equation}
v =a(c k_+ P - k_-).
\label{actinX}
\end{equation}
But application of the Boltzmann factor informs us that,
when the force on the membrane is $f$,
the probability that the gap is greater than $a$  is
\begin{equation}
P = e^{-fa/(k_B T)},
\end{equation}
so that
\begin{equation}
v =a(c k_+ e^{-fa/(k_B T)} - k_-).
\label{actinX}
\end{equation}
This is the force-velocity relationship for
an actin \cite{Kovar2004} or tubulin \cite{Dogterom1997} filament.
Although the load, $f$, is applied to the membrane, the velocity is constant. Therefore,
according to Newton's third Law, as the students know,
there can be no net force on the membrane. We may deduce that the load is balanced by an equal
and oppposite force, generated by the polymerization ratchet.

\section{Conclusions}
Three, interrelated biologically-relevant  examples of biased random walks were presented.
First, we presented a model
for DNA melting, modelled as DNA unzipping,
which provides a way to illustrate
the role of the Boltzmann factor in a venue well-known to the students. 
Second, we discussed
the activity of helicase motor proteins in unzipping double-stranded DNA,
for example, at the replication fork,
which is an example of a Brownian ratchet.
Finally, we treated force generation by actin polymerization, which
is another  Brownian ratchet, and for which
we can determine how the velocity of actin polymerization depends on actin
concentration and on load.
In each of these examples, building on an earlier coverage of biased random walks,
biology and pre-medical
students in an introductory physics sequence at Yale
were lead to the realization that a physics-based approach permits a deeper understanding of a familiar biological phenomenon.

\acknowledgments
I thank the Fall 2010 PHYS 170 class for their participation in PHYS 170,
and Sid Cahn, Stephen Eckel, Peter Koo, Lawrence Lee,
Andrew Mack, and Gennady Voronov for valuable discussions.
\bibliographystyle{unsrt}

\end{document}